\documentclass[aps,prl,twocolumn,preprintnumbers,nofootinbib,showpacs,floatfix]{revtex4}

\usepackage{graphicx}
\usepackage{amsmath}

\begin{document}

\title{Search for $W^{\prime}$ bosons through decays to boosted-top
  and boosted-bottom jets}

\author{Daniel~Duffty}
\email{dduffty@IIT.edu}
\affiliation{Department of Physics, Illinois Institute of Technology, Chicago, Illinois 60616-3793, USA}

\author{Zack~Sullivan}
\email{Zack.Sullivan@IIT.edu}
\affiliation{Department of Physics, Illinois Institute of Technology, Chicago, Illinois 60616-3793, USA}

\preprint{IIT-CAPP-13-03}

\begin{abstract}
  We propose a novel model-independent method to search for
  $W^{\prime}$ bosons at the Large Hadron Collider by looking at
  dijets where one jet is identified as a boosted-top jet and another
  jet is tagged as a boosted-bottom jet.  Performing a detector
  simulated signal and background study, we demonstrate that the reach
  in effective coupling $g^\prime$ is improved over existing analysis
  methods by up to factor of 2 for $W^{\prime}$ masses above 1.5~TeV,
  and the reach in mass is extended by 600~GeV to 2.6~TeV at an 8~TeV
  Large Hadron Collider.  We introduce a method for distinguishing
  high-energy $b$ jets from light-quark jets, and describe an
  important set of backgrounds to top-jets.  We propose a series of
  data-driven samples that might be used to measure the efficiencies
  for these new backgrounds in the LHC data.
\end{abstract}

\date{July 6, 2013}


\maketitle

\section{Introduction}
\label{sec:intro}

Model-independent searches for new charged vector currents, generally
called $W^\prime$ bosons, that decay to top and bottom quarks
\cite{Sullivan:2002jt} are important tools for investigating physics
beyond the standard model.  Following our recent theoretical
calculation of a model independent analysis at the Large Hadron
Collider (LHC) \cite{Duffty:2012rf}, the CMS \cite{:2012sc} and ATLAS
\cite{Aad:2012ej} Collaborations set strong bounds on generic
$W^\prime$ bosons as functions of effective coupling $g^\prime$.
While tighter mass bounds exist for special cases, such as pure
left-handed $W^\prime_L$ bosons using lepton transverse momentum
spectra \cite{CMSEXO}, the top quark final state exhibits a
resonant mass peak can be fully reconstructed \cite{Sullivan:2002jt}.
In this Letter we demonstrate the first use of boosted-top jets for
improving the reach in both $W^\prime$ mass and effective coupling
$g^\prime$.  In addition, we introduce a new boosted-bottom jet
algorithm that provides the key to extending this analysis beyond
existing measurements.

$W^\prime$ bosons appear in many classes of theories that stabilize
electroweak symmetry breaking.  High mass copies of $W$ boson
resonances are proposed in some theories \cite{Datta:2000gm};
right-handed counterparts to the left-handed standard model $W$ are
found in theories with a broken $SU(2)_L \times SU(2)_R$ symmetry
\cite{Pati:1974yy,Mohapatra:1974hk,Mohapatra:1974gc,Senjanovic:1975rk}. Other
theories, such as non-commuting extended technicolor
\cite{Chivukula:1995gu}, promote a heavy mass eigenstate.
$W^{\prime}$ bosons enter the Lagrangian with terms
of the form
\begin{equation}
\label{eq:L}
\mathcal{L}=\frac{g^\prime}{2\sqrt{2}} V^\prime_{ij} W^{\prime}_\mu \bar f^i \gamma^\mu 
(1\pm\gamma_5) f^j+\mathrm{H.c.}\,,
\end{equation}
which mirrors that of the standard model $W$ (without the lepton
sector if it is right-handed).  There could be left-right mixing, but
such mixing is constrained by $K$--$\overline{K}$ mixing
\cite{Groom:in}.  Theories have been proposed that would suppress this
mixing via orbifold breaking of the left-right symmetry
\cite{Mimura:2002te} or supersymmetric interactions
\cite{Cvetic:1983su}.

In searches for $W^\prime$ boson resonances between 1--3~TeV, existing
analyses \cite{Duffty:2012rf,:2012sc,Aad:2012ej} are limited by their
ability to reconstruct isolated decay products of the top quark ($t\to
bW\to bjj/bl\nu$).  As the mass of the $W^{\prime}$ boson increases,
relativistic angle compression causes the top quark decay products to
merge into a ``fat jet,'' distinguishable from normal jets only by its
substructure.  Several ``boosted top quark'' tagging algorithms have
been developed \cite{Kaplan:2008ie,CMS:2009lxa,Almeida:2008tp} to take
advantage of these kinematics.  So far these algorithms have shown
promise in Higgs plus top quark searches \cite{Plehn:2009rk} and
$Z^\prime \to t\bar t$ events \cite{Aad:2012raa}.  In this Letter, we
examine the effectiveness of boosted-top tags in improving the reach
to higher mass and smaller effective coupling.

In the first section of this Letter, we describe our simulation of the
$W^\prime$ signal and backgrounds in a dijet sample with a top tag.
We find naive theoretical estimates of boosted-top tag efficiencies
and mistag rates are too optimistic, and adopt the efficiencies
measured by the CMS Collaboration \cite{CMS:2009lxa}.  We also discuss
a class of backgrounds to top tagging that have been previously
ignored by experiments in hadronic channels, and place conservative
upper limits on these backgrounds for the signal we consider.  

We introduce a new ``boosted bottom quark tag'' in the second section
of this Letter to improve the reach in mass and couplings.  Utilizing
the kinematic features of TeV $b$-jets along with the muon decay
channel of $B$ hadrons, we find reasonable tagging efficiencies
and good fake rejection.  Since our analysis is at the level of a fast
detector simulation, we also recommend data samples where the
efficiencies can be extracted.

Having described models for signals and backgrounds using the new
boosted-top and bottom objects, we show that the reach in $W^\prime$
mass can be extended by 600~GeV in the same amount of data as used in
current limits.  More importantly, we find that limits on the
effective coupling $g^\prime$ can be improved by a factor of 2 over
the previous analysis techniques for masses above $1.5$~TeV.  Hence,
many perturbative models containing new charged-vector currents can be
excluded up to $2.6$~TeV.  We conclude this Letter with
suggestions for future research.

\section{Boosted-top tags and backgrounds}
\label {sec:toptag}

The process of interest is resonant $W^\prime$ production, with a
decay into a boosted-top jet and a boosted-bottom jet ($pp \to
W^{\prime} \to tb+X$) at the LHC.  The dominant backgrounds are
expected to come from dijet production ($jj$, $cj$, $c\bar c$, $bj$,
$b\bar b$, $bc$), where showering creates ``fat'' jets that fake boosted-top jets, and standard model (SM) production of single top
quarks in the $s$- and $t$-channel ($tb$, $tj$) and ditop ($t\bar t$),
which pass boosted-top jet tags.  In addition, we find a class of
``accidental'' backgrounds to boosted-top tags from $Wjj$ and $Zjj$
that we examine below.

We simulate events by feeding the output of a general $W^\prime$ model
\cite{ZhouSullivan} and backgrounds produced in \textsc{MadEvent}
\cite{Alwall:2007st} into \textsc{PYTHIA} \cite{Sjostrand:2006za} for
showering and hadronization.  We reconstruct events using an
ATLAS-like detector model in a modified version of the \textsc{PGS 4}
\cite{conway:2006pgs} detector simulation we have previously
calibrated to ATLAS data \cite{Duffty:2012rf}.  Cross sections are
scaled to next-to-leading order (NLO) using Ref.\ \cite{Duffty:2012rf}
for the $W^\prime$ signal, and \textsc{MCFM} \cite{Campbell:2010ff}
for backgrounds.

In order to make contact with experiment, we begin with a
Cambridge-Aachen implementation of boosted-top tagging studied by the
CMS Collaboration \cite{CMS:2009lxa}.  The CMS procedure is quite
involved, but it roughly corresponds to finding a large ($\Delta
R<0.8$) hard jet whose mass loosely reconstructs to a top-quark mass
between 100--250~GeV.  In addition, the jet must decompose into
three or four ``subjets,'' of which a pair must combine to produce a
$W$-like mass ($> 50$~GeV).  Implementations of this top tagging
algorithm are extremely sensitive to details of the
detector simulation, such as non-uniform calorimeter segmentation.  In
implementing the CMS algorithm, we initially found top-jet
acceptances $\sim 10\%$ higher than observed by CMS, and fake rates
from light jets nearly an order-of-magnitude lower than observed by
CMS.

To reduce sensitivity to our detector model and provide a more
realistic comparison between $W^\prime$ reconstruction using top jets
and analyses where top decay products can be isolated
\cite{Duffty:2012rf}, we fit transverse momentum ($p_T$) dependent
tagging efficiencies for top-jets and fakes rates provided by CMS in
Figs.\ 13 and 14, respectively, of Ref. \cite{CMS:2009lxa} and apply
them to our reconstructed jets.\footnote{Initial efficiency estimates
  by the ATLAS Collaboration using an anti-$k_T$ algorithm are
  quantitatively consistent with the fits we use here
  \protect\cite{privateJeremy}, suggesting low sensitivity to details
  of the algorithm.}  While the CMS top-jet acceptance is
asymptotically 45\% at large $p_T$, our results are driven primarily
by the fake rate, which ranges from 1--4\% over the $p_T$ range we
reconstruct.  This large fake rate would cause $jj$ to completely
dominate the signal if not for the new boosted-bottom tag introduced
below.

\begin{figure}
\includegraphics[width=0.9\columnwidth]{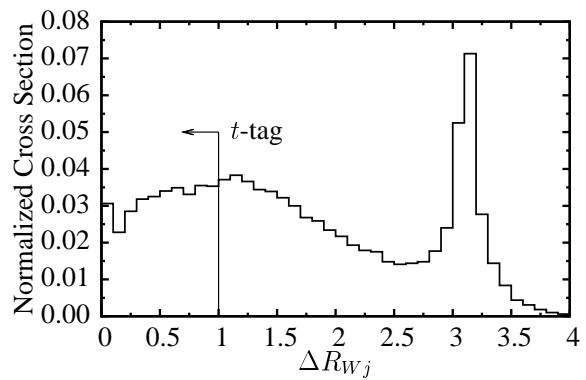}
\caption{$\Delta R$ between the $W$ and the nearest jet in $Wjj$
  events.  Events with $\Delta R < 1$ are candidates for being tagged
  as a top jet, provided they fit a loose top-mass
  cut. \label{fig:R_Wjj}}
\end{figure}

Nothing in boosted-top jet algorithms requires the subjets to come
from top quarks or showered light-quark jets (fakes).  An important
class of backgrounds to top tagging comes from
$Wjj/Wbj$ or $Zjj$ events, where one jet fluctuates close
enough to a hadronically decaying $W$ or $Z$ to pass the boosted-top
algorithm.  We model these contributions by using the detector
simulation above to look for three jets in a cone of $\Delta R<1$ and
applying a cut on the invariant mass of the three jets of
$100<M_{jjj}<250$ GeV.  In Fig.\ \ref{fig:R_Wjj} we see a significant
fraction of $Wjj$ events pass the cut in $\Delta R$, and we show below
that this background is larger than either real standard model top or
bottom quark processes after all cuts.
Similar backgrounds in semi-leptonic boosted-top decays have been
examined by considering muon proximity to jets
\cite{Rehermann:2010vq}; and a parameter $z_{\mu}$ has been proposed
\cite{Thaler:2008ju} to suppress this non-top background.  It is
possible similar methods could be developed to suppress $Wj+X$ contributions
to hadronic boosted-top decays.

A significant challenge to estimating the backgrounds from $Wjj$ or
$Zjj$ comes from instabilities of the perturbative calculation in the
highly restricted phase space when a jet is near the $W$.  In order to
place upper bounds on these contributions we estimate a theoretical
error of 50\% in the acceptance using two methods: First we look at
increases in acceptance between leading order and NLO predicted by
MCFM \cite{Campbell:2010ff} due to higher order hard radiation near
the $W$.  In addition, we examine increases in acceptance between
parton level and showering after \textsc{PYTHIA} and reconstruction.
Both methods indicate at least 30\% increases in boosted-top
acceptance of $Wjj$, $Wbj$, and $Zjj$.  Hence, we include an
additional $K$-factor of $1.5$ times the NLO cross section when
calculating these backgrounds.  Fortunately, these backgrounds are
suppressed by other cuts, but future boosted-top studies should
account for them.

\section{Boosted-bottom jet tagging}
\label {sec:btag}

In a $W^\prime\to tb$ final state, the jet containing the $B$ hadron
which recoils against the top jet is typically the highest energy jet
in the event.  Historically, the flavor content of the $b$-jet has
been ignored
\cite{Sullivan:2002jt,Acosta:2002nu,Aaltonen:2009qu,Abazov:2006aj,Abazov:2011xs,:2012sc,Aad:2012ej}
because modern secondary vertex tagging is difficult above a few
hundred GeV.  The decay products of $B$ hadrons with relativistic
$\gamma=100$--$300$ are boosted to be approximately collinear, and
hence tracking generally points back to the interaction vertex within
the resolution of the detectors.  In order to overcome the dijet
backgrounds due to light-quark jets, it is necessary to identify these
$b$ jets.

\begin{figure}
\includegraphics[width=0.9\columnwidth]{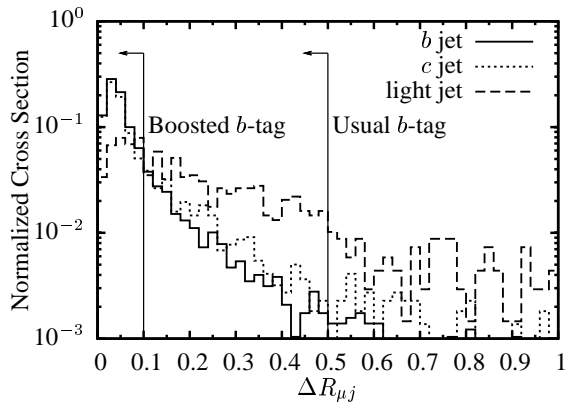}
\caption{$\Delta R$ separation between muons and the centroid of their
  associated jets in $bb$, $cc$, and $jj$ events for jets with $E_T >
  400$ GeV.  Muons from $b$ decays are more central than those from
  light quark jets.\label{fig:DelRmuon}}
\end{figure}

We propose a new ``boosted-bottom tagging'' algorithm that adds a
kinematic twist to the old idea of $b$-tagging by observation of a
muon inside of a jet.  Beginning with data at the Fermilab Tevatron,
muon-tagging of $b$ jets was relegated to enhancing the efficiency of
secondary vertex tags.  Typically, muons found within a 30 degree cone
($\Delta R_{\mu j}\sim 0.5$) of the centroids of jets provide 10--15\%
of the $b$-tagging efficiency \cite{conway:2006pgs}.  The fake rate
for these muons is dominated by long-lived kaon and charged pion
decays inside of the jets.  As can be seen in Fig.\
\ref{fig:DelRmuon}, however, in jets with transverse energies of
$E_T=500$--1500~GeV almost all of the muons from bottom or charm decays
are boosted to within $\Delta R_{\mu j}< 0.1$, while muons from
light-quark or gluon initiated jets have a broader spectrum in $\Delta
R_{\mu j}$.  This contrasts with usual $b$-tagging, where some
fraction of muons is either soft or at wide angles to the $b$
direction.  Hence, our method for tagging boosted-bottom jets is to
look for hits in a muon chamber within $\Delta R_{\mu j}< 0.1$.

In addition to the smaller angle between the muon from $b$ decay and
the jet, the transverse momentum of the muons from these
boosted-bottom jets tends to be significantly harder that that of the
light jets.  We place a cut of $p_{T}>20$~GeV on the muon in order to
aid reconstruction.  This value corresponds to roughly the minimum
possible muon energy in the LAB frame for these boosted kinematics.
As luminosities increase at the LHC this threshold could be made
proportional to the jet energy with a modest loss of acceptance.  We
extract transverse-energy $E_T$-dependent efficiencies for $b$, $c$,
and light jet tags and list a few representative points in Tab.\
\ref{tab:btagrate}.  The ratio of $b$-tagging efficiency to fake rates
is similar to what is currently achieved by secondary vertex tagging.
The drop in efficiency below 400~GeV is an artifact of using a fixed
angular cut to maintain $b$ jet purity.

\begingroup
\begin{table}[!htb]
  \caption{Boosted-bottom jet efficiencies using a muon tag with
    $p_{T\mu}>20$~GeV and $\Delta R_{\mu j}<0.1$ for $b$ jets, $c$ jets,
    and light jets $j$ as a function of jet $E_T$.\label{tab:btagrate}}
\begin{ruledtabular}
\begin{tabular}{cccc}
Type & $E_{T j}=100$ GeV & $400$ GeV & $1000$ GeV \\
\hline
$b$ & 4.8$\%$ & 11.8$\%$ & 15.0$\%$ \\
$c$ & 2.1$\%$ & 5.5$\%$ & 7.5$\%$ \\
$j$ & 0.1$\%$ & 0.4$\%$ & 0.6$\%$ \\
\end{tabular}
\end{ruledtabular}
\end{table}
\endgroup

For this study we use boosted-bottom tag and fake rates estimated via
Monte Carlo, but it is important to confirm these rates in data.
Ideally one would want to examine a $b\bar b$ final state where the
energies of the $b$ jets are near to the final boosted energies,
however secondary vertexing is limited near 1~TeV.  One promising
sample would be $bbj$, where one boosted-$b$ jet recoils against a
softer $b$ jet that is secondary vertex- (or muon-) tagged, and a
light-quark or gluon jet.  Based on our analyses there may be a few
hundred boosted-$b$ jets near 1~TeV in the 20~fb$^{-1}$ sample already
collected at the LHC.  The fake rate could be obtained by comparing to
the much larger high-$E_T$ dijet sample.  Another possible sample with
some reach for boosted $b$-tags may be the high-$E_T$ tail of the
dilepton channel of $t\bar t$ production, where the boosted-$b$ jet
contains a muon, and the second lepton is isolated enough to be
distinguished.  Given the size of the background from $t\bar t$ to the
standard model-independent analysis, there may be enough isolated
events to get initial estimates of the efficiencies.

\section{$W^{\prime}$ search with boosted-$t$ and $b$ jets}
\label{sec:wp}

We now turn to using boosted-top jet tags and boosted-bottom jet
tags to examine their reach in probing for $pp\to W^\prime\to tb$
in the dijet sample for masses above 1~TeV.  Unlike the analysis of
Ref.\ \cite{Duffty:2012rf}, angular correlations of the top decay
products are hidden by the kinematics of the two body decay.  While
interference between left-handed $W^\prime_L$ bosons with the standard
model affects the total production cross section, we reconstruct
events in the resonance region where the interference is negligible.
Hence, we are left with tagging as one of the few handles in the
analysis.

We begin with a sample of reconstructed central high energy dijets
with $E_{Tj_1}> 0.38M_{W^\prime}$, $E_{Tj_2}> 0.1M_{W^\prime}$,
$|\eta_j| < 1.5$, and $\Delta R_{jj} > 0.4$.  We demand the
pseudorapidity between the two jets is also small $|\eta_{j_1} -
\eta_{j_2}| < 1.5$ in order to suppress large angle contributions to
the dijet invariant mass $M_{jj}$.  For each $W^\prime$ mass we
reconstruct a dijet invariant mass in $0.95 M_{W^\prime} < M_{jj} <
1.05 M_{W^\prime}$.  We set 95\% confidence level (C.L.) exclusion
limits after cuts.

In Tab.\ \ref{tab:tageffects} we list the number of signal $S$ and
background $B$ events expected for a 2~TeV $W^\prime$ with
$g^\prime/g_{SM}=1$ and 20~fb$^{-1}$ at a 8~TeV LHC.  After acceptance
cuts, direct dijet production in QCD is 1000 times the signal.  Adding
a boosted-top tag improves $S/B$ to $1/33$, but the accidental
backgrounds $Wjj/Wbj$ and $Zjj$ are larger than standard model $t\bar
t$, single top, and comparable or larger than heavy-flavor dijets.
Hence, future analyses of boosted-top jets should not ignore these
accidentals.

\begingroup
\begin{table}[!htb]
  \caption{Number of $W^\prime\to tb$ signal and background events predicted
    for a 2~TeV $W^\prime$ boson in 20~fb$^{-1}$ of integrated luminosity at an
    8~TeV LHC.  Events are listed after acceptance cuts, with a boosted-top tag
    ($+t$-tag), and adding a boosted-bottom tag ($+b$-tag).
    Accidental backgrounds $Wjj/Wbj$ and $Zjj$ are not present before top
    tagging.\label{tab:tageffects}}
\begin{ruledtabular}
\begin{tabular}{lrrr}
Events & After acceptance
& $+$ $t$-tag
& $+$ $b$-tag \\
\hline
$W^\prime$ signal & 702 & 316 & 47 \\
$jj$ & 735000 & 11000 & 66  \\
$c\bar c$, $cj$ & 19300 & 289 & 12  \\
$b\bar b$, $bj$, $bc$ & 10800 & 162 & 14  \\
$t\bar t$, $tj$, $tb$ & 51 & 25 & 4  \\
$Wbj$, $Wjj$, $Zjj$ & * & 93 & 1 \\
\end{tabular}
\end{ruledtabular}
\end{table}
\endgroup

\begin{figure}
\includegraphics[width=0.9\columnwidth]{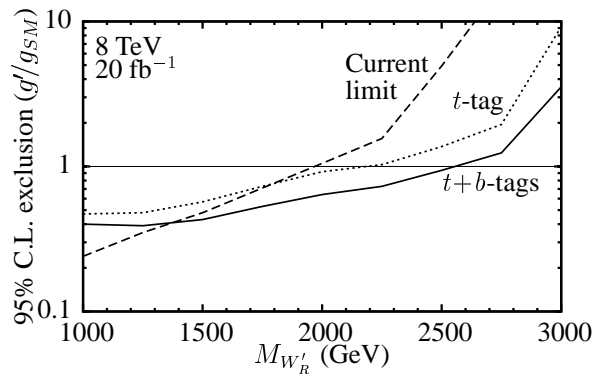}
\caption{95\% C.L.\ limit on the effective coupling $g^\prime_R$ relative
to $g_{SM}$ as a function of right-handed $W^\prime_R$ mass. Curves show the
reach from current resolved-top quark analyses (dashed), the boosted-top
analysis (dotted), and after adding a boosted-bottom tag (solid).
\label{fig:couplingR}}
\end{figure}

\begin{figure}
\includegraphics[width=0.9\columnwidth]{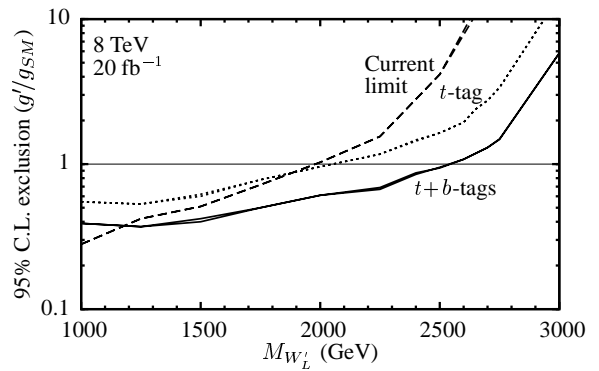}
\caption{95\% C.L.\ limit on the effective coupling $g^\prime_L$ relative
to $g_{SM}$ as a function of left-handed $W^\prime_L$ mass. 
Constructive (lower lines) and destructive (higher lines)
interference with the standard model are nearly indistinguishable.
\label{fig:couplingL}}
\end{figure}

In Figs.\ \ref{fig:couplingR} and \ref{fig:couplingL} we see that
using a boosted-top tag improves the 95\% C.L.\ exclusion limit on
$g^\prime/g_{SM}$ slightly for $M_{W^\prime} > 2$~TeV if
$g^\prime/g_{SM}>1$.  The region $g^\prime/g_{SM}>1$ is difficult to
measure as the width of the $W^\prime$ resonance grows rapidly with
coupling ($\Gamma \sim 2.5\%\times M_{W^\prime}(g^\prime/g_{SM})^2$).
We treat the full width effects using Refs.\
\cite{Duffty:2012rf,ZhouSullivan} since naive coupling scaling
\cite{Sullivan:2002jt,Duffty:2012rf} is only valid for
$g^\prime/g_{SM}<1$.  Small differences between right- and left-handed
$W^\prime$ are due to different $tb$ branching fractions.  In Fig.\
\ref{fig:couplingL} we show both positive or negative interference
between $W^\prime_L$ and the standard model.  As expected, the
interference is tiny in the resonance region.

In order to reach the full potential of this channel we next apply the
new boosted-bottom tag.  We see in Tab.\ \ref{tab:tageffects} the
sample purity $S/B$ improves to $1/2$, and the significance
$S/\sqrt{B}$ improves by 50\% for $M_{W^\prime}=2$~TeV.  The
boosted-bottom tag increases the reach in $W^\prime$ mass with
$g^\prime/g_{SM}=1$ by 600~GeV.  More importantly, the limit in
$g^\prime/g_{SM}$ improves by nearly a factor of 2 over existing
approaches for masses above 1.5~TeV, and remains perturbative to
3~TeV.  Hence, models with slightly greater than standard model
coupling, like orbifolded left-right breaking \cite{Mimura:2002te},
can be ruled out for masses below 3~TeV.  Combined with previous
analyses below 1.5~TeV, a wide range of perturbative models can be
excluded for masses below 2.6~TeV.

\section{Conclusions}
\label{sec:concl}

In this Letter we present the first use of boosted-top tags and
introduce boosted-bottom tags in the search for new charged vector
currents, called $W^\prime$ bosons, by looking at the $tb$ dijet final
state.  We present a quantitative estimate of ``accidental''
backgrounds to boosted-top jet measurements which are comparable to or
larger than all backgrounds other than direct dijet $jj$.  We
implement a detector simulated analysis whose efficiencies are fit to
data, and find that boosted-top jets provide modest improvement on
existing analyses of the $tb$ final state which use isolated leptons
and jets from top-quark decay.

The key to improving the experimental reach in $W^\prime$ mass by 600
GeV, and model-independent coupling $g^\prime$ by roughly a factor of
2, is the use of a new ``boosted-bottom jet tag.''  This
boosted-bottom tag should be useful beyond the signal presented here.
In particular, studies of $Z^\prime$ production to boosted-top can be
supplemented by studies of $Z^\prime\to b\bar b$.  The smaller
acceptance may be compensated by the greater background rejection
from boosted-bottom tagging, and enhanced coupling to down-type quarks
than to up-type quarks.  Any channel with bottom jets in the final
state is now open to direct observation.

Note added: After submission of this Letter, studies of pileup at a
14~TeV high luminosity LHC suggest boosted top tags will remain efficient
\cite{Calkins:2013ega}.

\begin{acknowledgments}
  This work is supported by the U.S.\ Department of Energy under
  Contract No.\ DE-SC0008347.
\end{acknowledgments}



\begin{thebibliography}{99}

\bibitem{Sullivan:2002jt} 
Zack~Sullivan,
Phys.\ Rev.\ D {\bf 66}, 075011 (2002).

\bibitem{Duffty:2012rf} 
Daniel~Duffty and Zack~Sullivan,
Phys.\ Rev.\ D {\bf 86}, 075018 (2012).

\bibitem{:2012sc}
S.~Chatrchyan {\it et al.}  (CMS Collaboration),
Phys.\ Lett.\ B {\bf 718}, 1229 (2013).

\bibitem{Aad:2012ej} 
  G.~Aad {\it et al.}  (ATLAS Collaboration),
Phys.\ Rev.\ Lett.\  {\bf 109}, 081801 (2012).

\bibitem{CMSEXO}
CMS Collaboration, CMS PAS EXO-12-060.

\bibitem{Datta:2000gm}
A.~Datta, P.~J.~O'Donnell, Z.~H.~Lin, X.~Zhang and T.~Huang,
Phys.\ Lett.\ B {\bf 483}, 203 (2000).

\bibitem{Pati:1974yy}
J.~C.~Pati and A.~Salam,
Phys.\ Rev.\ D {\bf 10}, 275 (1974).

\bibitem{Mohapatra:1974hk}
R.~N.~Mohapatra and J.~C.~Pati,
Phys.\ Rev.\ D {\bf 11}, 566 (1975).

\bibitem{Mohapatra:1974gc}
R.~N.~Mohapatra and J.~C.~Pati,
Phys.\ Rev.\ D {\bf 11}, 2558 (1975).

\bibitem{Senjanovic:1975rk}
G.~Senjanovic and R.~N.~Mohapatra,
Phys.\ Rev.\ D {\bf 12}, 1502 (1975).

\bibitem{Chivukula:1995gu}
R.~S.~Chivukula, E.~H.~Simmons and J.~Terning,
Phys.\ Rev.\ D {\bf 53}, 5258 (1996).

\bibitem{Groom:in}
D.~E.~Groom {\it et al.}  (Particle Data Group Collaboration),
Eur.\ Phys.\ J.\ C {\bf 15}, 1 (2000).

\bibitem{Mimura:2002te}
Y.~Mimura and S.~Nandi,
Phys.\ Lett.\ B {\bf 538}, 406 (2002).

\bibitem{Cvetic:1983su}
M.~Cvetic and J.~C.~Pati,
Phys.\ Lett.\ B {\bf 135}, 57 (1984).

\bibitem{Kaplan:2008ie} 
  D.~E.~Kaplan, K.~Rehermann, M.~D.~Schwartz, and B.~Tweedie,
  Phys.\ Rev.\ Lett.\  {\bf 101}, 142001 (2008).

\bibitem{CMS:2009lxa}
  CMS Collaboration,
  CMS-PAS-JME-09-001.

\bibitem{Almeida:2008tp} 
  L.~G.~Almeida, S.~J.~Lee, G.~Perez, I.~Sung, and J.~Virzi,
  Phys.\ Rev.\ D {\bf 79}, 074012 (2009).

\bibitem{Plehn:2009rk} 
  T.~Plehn, G.~P.~Salam, and M.~Spannowsky,
  Phys.\ Rev.\ Lett.\  {\bf 104}, 111801 (2010).

\bibitem{Aad:2012raa} 
G.~Aad {\it et al.}  (ATLAS Collaboration),
J.\ High Energy Phys.\ {\bf 1301}, 116 (2013).

\bibitem{ZhouSullivan} Yaofu~Zhou and Zack~Sullivan, 
``Modeling $W^{\prime}$ bosons for use with model independent studies,''
paper in production.

\bibitem{Alwall:2007st}
J.~Alwall {\it et al.},
J.\ High Energy Phys.\ {\bf 0709}, 028 (2007).

\bibitem{Sjostrand:2006za} 
T.~Sjostrand, S.~Mrenna, and P.~Z.~Skands,
J.\ High Energy Phys.\ {\bf 0605}, 026 (2006).

\bibitem{conway:2006pgs} 
J.\ Conway  {\it et al.},
\url{http://www.physics.ucdavis.edu/~conway/research/software/pgs/pgs.html.}

\bibitem{Campbell:2010ff} 
J.~M.~Campbell and R.~K.~Ellis,
Nucl.\ Phys.\ Proc.\ Suppl.\  {\bf 205-206}, 10 (2010).

\bibitem{privateJeremy}
J.\ Love, private communication.

\bibitem{Rehermann:2010vq} 
K.~Rehermann and B.~Tweedie,
J.\ High Energy Phys.\ {\bf 1103}, 059 (2011).

\bibitem{Thaler:2008ju} 
J.~Thaler and L.~-T.~Wang,
J.\ High Energy Phys.\ {\bf 0807}, 092 (2008).

\bibitem{Acosta:2002nu} 
D.~Acosta {\it et al.}  (CDF Collaboration),
Phys.\ Rev.\ Lett.\  {\bf 90}, 081802 (2003).

\bibitem{Aaltonen:2009qu} 
T.~Aaltonen {\it et al.} (CDF Collaboration),
Phys.\ Rev.\ Lett.\  {\bf 103}, 041801 (2009).

\bibitem{Abazov:2006aj} 
V.~M.~Abazov {\it et al.}  (D0 Collaboration),
Phys.\ Lett.\ B {\bf 641}, 423 (2006).

\bibitem{Abazov:2011xs}
V.~M.~Abazov {\it et al.} (D0 Collaboration),
Phys.\ Lett.\ B {\bf 699}, 145 (2011).

\bibitem{Calkins:2013ega} 
R.~Calkins {\it et al.},
in ``Proceedings of the Community Summer Study 2013 --- Snowmass on
the Mississippi,'' SNOW13-00075, arXiv:1307.6908 [hep-ph].


\end{thebibliography}
\end{document}